\documentclass[prd, 10pt,aps,nofootinbib,floatfix,notitlepage,twocolumn,superscriptaddress]{revtex4-2}

\usepackage{float}
\usepackage[dvipsnames]{xcolor}
\usepackage{graphicx}
\usepackage{dcolumn}
\usepackage{bm}
\usepackage{amsmath}
\usepackage[colorlinks]{hyperref}
\usepackage{xspace}
\usepackage{subfigure}
\usepackage{verbatim}
\usepackage{amssymb}
\usepackage{ulem}
\usepackage{multirow}
\usepackage[capitalise]{cleveref}
\crefname{section}{Sec.}{Secs.}
\crefname{appendix}{App.}{App.}

\usepackage{tabularx}
\newcolumntype{L}{>{\raggedright\arraybackslash}X}
\newcolumntype{C}{>{\centering\arraybackslash}X}
\newcolumntype{R}{>{\raggedleft\arraybackslash}X}

\hypersetup{
    linkcolor=NavyBlue,
    citecolor=NavyBlue,
    urlcolor=NavyBlue,
    allcolors=NavyBlue
}

\newcommand{\Planck}{\textit{Planck}}
\newcommand{\hhat}[1]{\hat{\hat{#1}}}
\renewcommand{\d}{\mathrm{d}}
\newcommand{\LCDM}{$\Lambda$CDM\xspace}

\begin{document}
\title{Bayesian and frequentist perspectives agree on dynamical dark energy}

\author{Laura Herold}
\email{lherold@jhu.edu}
\affiliation{William H. Miller III Department of Physics and Astronomy, Johns Hopkins University, 3400 North Charles Street, Baltimore, MD 21218, USA}
\author{Tanvi Karwal}
\email{karwal@uchicago.edu}
\affiliation{Kavli Institute for Cosmological Physics, Enrico Fermi Institute, and Department of Astronomy \& Astrophysics, University of Chicago, Chicago, IL 60637, USA}

\begin{abstract}
    Baryon acoustic oscillation data from the Dark Energy Spectroscopic Instrument (DESI) show evidence of a deviation from a cosmological constant $\Lambda$ within a Bayesian analysis. 
    In this work, we validate that frequentist constraints from profile likelihoods on the Chevallier-Polarski-Linder parameters $w_0$, $w_a$ are in excellent agreement with the Bayesian constraints when combining with \textit{Planck} cosmic microwave background, \textit{Planck} and Atacama Cosmology Telescope lensing, and either Pantheon+ or Dark Energy Survey Y5 supernova data. Further, we assess which datasets drive these constraints by considering the contributions to the $\chi^2$  from the individual datasets. For profile likelihoods of the matter fraction $\Omega_\mathrm{m}$, such an investigation shows internal inconsistencies when assuming $\Lambda$, which are resolved when assuming a $w_0w_a$ dark-energy model. We infer the equations of state $w(z)$ at the pivot redshifts, supporting previous interpretations that current data appear to be more sensitive to the derivative of $w(z)$ rather than a mean offset from $\Lambda$. Thus our frequentist analysis corroborates previous findings on dynamical dark energy.
\end{abstract}

\maketitle


\section{Introduction}
\label{sec:intro}

Through precision spectroscopic measurements of baryon acoustic oscillations (BAOs), the Dark Energy Spectroscopic Instrument (DESI) has recently reported hints of dynamical dark energy (DE) \cite{DESI:2024mwx, DESI:2024kob,DESI:2025zgx, DESI:2025gwf}. While this tentative preference for dynamical DE exists across various phenomenological and fundamental models \cite{Berghaus:2024kra, DESI:2025wyn,DESI:2025fii, Yang:2025mws}, constraints are often presented in the Chevallier-Polarski-Linder (CPL, \cite{Chevallier:2000qy, Linder:2002et}) parameterization of the equation of state of DE. Also commonly known as the ``$w_0w_a$'' model of dynamical DE, it ascribes the following redshift-evolution to the equation of state $w$ of DE 
\begin{equation}
	w(a)  = w_0 + w_a (1-a) \,,
	\label{eq:CPL}
\end{equation}
where $a$ is the scale factor, and $w_0$ and $w_a$ are constants. 

Combining DESI Data Release (DR) 2 BAO data with cosmic microwave background (CMB) data, the DESI collaboration reported (Eq.~25 of \cite{DESI:2025zgx})
\begin{align}
    w_0 &= -0.42 \pm 0.21 \,\, {\rm and} \nonumber \\
    w_a &= -1.75 \pm 0.58 \,, \nonumber
\end{align}
indicating an exclusion of a cosmological constant $\Lambda$ with constant equation of state $w=-1$ at approximately $3\sigma$. Including data from uncalibrated type-Ia supernovae (SNe) further increases the significance to $3-4\sigma$ \cite{DESI:2025zgx,DESI:2025fii}. While the inclusion of DESI BAO data yields the highest levels of significance, DESI-independent datasets also show tentative preference for a deviation from the $\Lambda$ cold dark matter (\LCDM) model \cite{Brout:2022vxf, DES:2025bxy, Giare:2025pzu, Scherer:2025esj}.

These $68\%$ credible intervals are based on Bayesian posteriors obtained with Markov-chain Monte Carlo (MCMC) methods. Given the relevance of these results, in this work, we verify whether frequentist confidence intervals agree with Bayesian credible intervals for the parameters of the CPL dynamical-DE model (see also \cite{Tang:2024lmo,Staicova:2025tbi} which explore marginalized and profiled constraints from DESI BAO data). While the DESI collaboration has already quantified deviations from $\Lambda$CDM using frequentist $\Delta\chi^2$, we extend their work by calculating frequentist confidence intervals.

Frequentist parameter-inference methods based on profile likelihoods are sparingly applied in cosmology, but they have recently received increased attention for their ability to provide complementary information to Bayesian approaches. While Bayesian credible intervals rely on a choice of prior on all model parameters and handle nuisance parameters through integration (``marginalization''), frequentist confidence intervals do not necessarily require a prior and usually handle nuisance parameters through maximization. 
This difference in prior-selection and nuisance-parameter treatment becomes particularly important for models that are poorly constrained by the available data, such as when the likelihood is flat or deviates significantly from a Gaussian distribution. In a Bayesian analysis, this can lead to prior(-volume) effects (see e.g.\ \cite{Feeney:2013wp, Gariazzo:2018pei, Smith:2020rxx,Herold:2021ksg, Herold:2022iib, Gariazzo:2022ahe, Poulin:2023lkg,Holm:2023laa,Karwal:2024qpt} for examples in a cosmology context). These complementary approaches converge in the limit of large datasets. 

\begin{figure*}
    \centering
    \includegraphics[width=\linewidth]{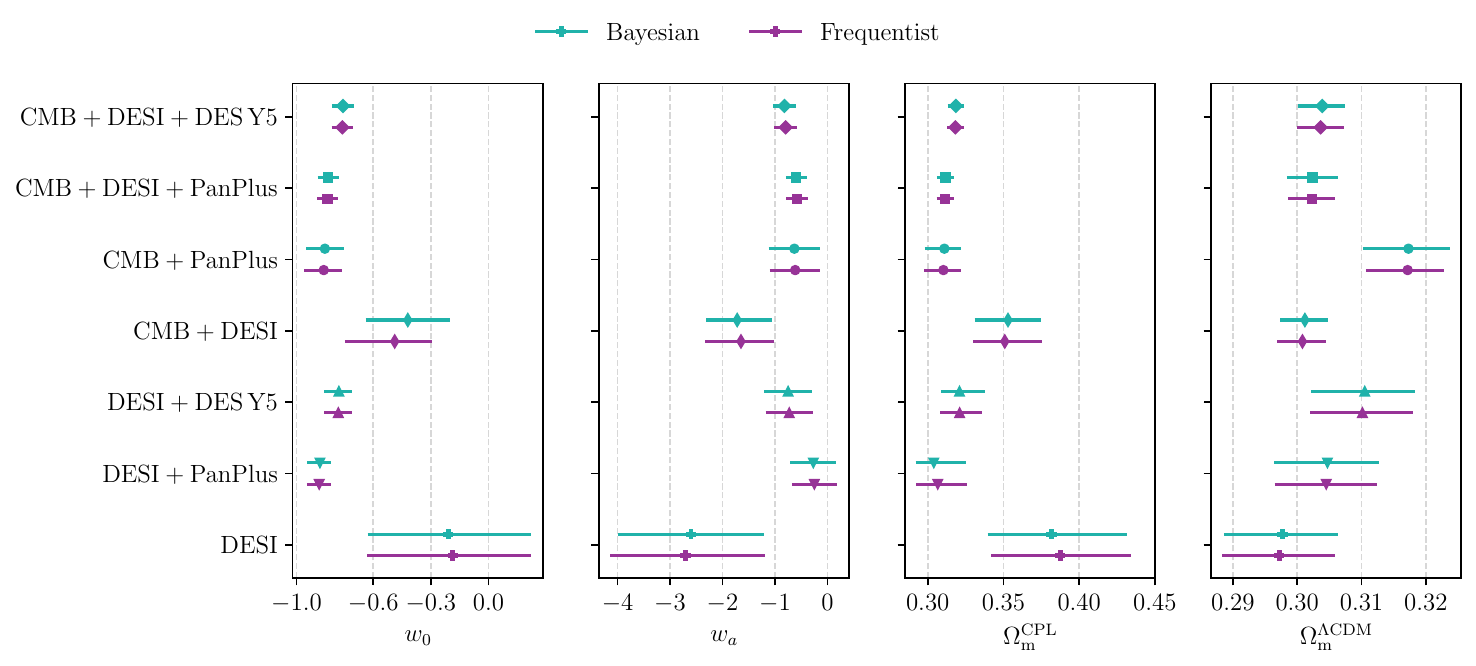}
    \caption{
    Comparison of constraints from Bayesian (mean $\pm 1\sigma$) and frequentist statistical methods (MLE $\pm 1\sigma$) for $w_0 \,, w_a$ and $\Omega_{\rm m}^{\rm CPL}$ within the CPL model and $\Omega_{\rm m}^{\Lambda {\rm CDM}}$ within the $\Lambda$CDM model for various dataset combinations. CMB denotes \Planck\ CMB primaries and lensing from \Planck\ and ACT combined.
    Bayesian posteriors for each combination (except for DESI alone) constrain CPL and \LCDM parameters well within their prior bounds, and the two statistical methods are in excellent agreement. 
    }
    \label{fig:whisker}
\end{figure*}

In this work we supplement the Bayesian constraints in the literature with frequentist constraints obtained with profile likelihoods. Our goal is to verify whether the frequentist constraints on the CPL parameters $w_0\,,w_a$ agree with their Bayesian counterparts. The results of this paper are summarized in \cref{fig:whisker} and show excellent agreement between the two approaches.

Although the CPL model is a parametric model with no underlying microphysics (see e.g.\ \cite{Wolf:2023uno, Nesseris:2025lke} for discussions), it remains the most broadly studied dynamical-DE model. Moreover, the choice of priors on the CPL parameters has been a subject of discussion in the literature \cite{Cortes:2024lgw, Patel:2024odo, Efstathiou:2025tie}. We hence focus specifically on CPL in this study. We further discuss the impact of different values of the fractional matter energy density $\Omega_\mathrm{m}$ preferred by CMB, BAO and SN data and the choice of pivot scale of the CPL parameterization. 

This paper is organized as follows. In \cref{sec:data_methods}, we detail the datasets and our methodology. \cref{sec:results} presents the results of our frequentist analysis, and \cref{sec:discussion} describes the impact of the preferred $\Omega_\mathrm{m}$ values and the pivot scale of the CPL parameterization. We conclude in \cref{sec:conclusions}.


\section{Datasets and methodology}
\label{sec:data_methods}

We use the latest DESI DR2 BAO sample \cite{DESI:2025zpo, DESI:2025zgx} (referred to as DESI). While the public DESI BAO likelihood is compatible with the \texttt{Cobaya} sampler \cite{Torrado:2020dgo}. We adapt this likelihood to be compatible with the \texttt{MontePython} sampler \cite{Audren:2012wb, Brinckmann:2018cvx}\footnote{We make the DESI \texttt{MontePython} likelihood and the notebooks to reproduce some of the plots in this paper available at \url{https://github.com/LauraHerold/MontePython_desilike}.}.  

Following the convention of the DESI collaboration, we combine their BAO data with primary CMB data from \Planck\ PR3 (TT, TE, EE power spectra) \cite{Planck:2019nip, Planck:2018vyg} and CMB lensing power spectra from the Atacama Cosmology Telescope (ACT) DR6  \cite{ACT:2023kun} combined with \Planck\ PR4 \cite{Carron:2022eyg} (jointly referred to as CMB). The inclusion of the recent CMB lensing data requires an increase in precision settings recommended by the ACT collaboration (see App.\ A in \cite{ACT:2025tim}), incorporated by the DESI collaboration data we compare to.

Additionally, we consider uncalibrated type-Ia SNe from the Pantheon+ \cite{Scolnic:2021amr, Brout:2022vxf} and the Dark Energy Survey (DES) Y5 samples \cite{DES:2024jxu, DES:2024upw}. We do not consider the Union SN sample \cite{Rubin:2023ovl} in this work since we do not expect a qualitatively different behavior from the Pantheon+ and DES Y5 samples\footnote{We adapt the DES Y5 likelihood to be compatible with \texttt{MontePython} and make it available at \url{https://github.com/tkarwal/cosmo_likelihoods}. Note that this is the DES Y5 sample before the Dovekie update \cite{DES:2025sig}.
}. 

We use the Boltzmann solver \texttt{CLASS} \cite{Lesgourgues:2011re, Blas:2011rf} for linear cosmological predictions and \texttt{HMCode} for non-linear corrections \cite{Mead:2020vgs}. The total sum $M_\mathrm{tot}$ of neutrino masses is fixed to the minimum allowed mass $M_\mathrm{tot} = 0.06\, \mathrm{eV}$ in the normal hierarchy, assuming the degenerate-mass approximation where each neutrino species carries $M_\mathrm{tot}/3$. This neutrino hierarchy is slightly different from the one assumed in \cite{DESI:2025zgx}, which assumes one massive and two massless neutrinos for the analyses with fixed $M_\mathrm{tot}$ (see e.g.\ \cite{Lesgourgues:2006nd, Giusarma:2016phn, Vagnozzi:2017ovm, Archidiacono:2020dvx, Herold:2024nvk} for arguments supporting the degenerate-mass approximation).

The Bayesian posteriors are obtained with \texttt{MontePython} \cite{Audren:2012wb, Brinckmann:2018cvx}  and illustrated with \texttt{GetDist} \cite{Lewis:2019xzd}. We assume flat priors $w_0 \in [-3, 1]$ and $w_0+w_a \in [-5, 0]$, which are similar to the ones assumed by \cite{DESI:2025zgx}, i.e.\ $w_0 \in [-3, 1]$, $w_a \in [-3, 2]$, $w_0+w_a < 0$. We consider the chains converged when the Gelman-Rubin criterion reaches $R-1 < 0.05$. 

The frequentist profile likelihoods are obtained with \texttt{pinc} \cite{Herold:2024enb}\footnote{Publicly available at \url{https://github.com/LauraHerold/pinc}.}, which uses simulated-annealing-based minimization interfaced with \texttt{MontePython} and crosschecked with \texttt{Procoli} \cite{Karwal:2024qpt}, a similar code. In a nutshell, a profile likelihood over the parameter $\theta$ of interest (here $\theta \in [w_0, w_a, \Omega_\mathrm{m}]$) is computed as
\begin{equation}
    \label{eq:PL}
    \Delta\chi^2 (\theta) = -2\log \left(\frac{\mathcal{L}(\theta,\hhat{\boldsymbol{\nu}})}{\mathcal{L}(\hat{\theta},\hat{\boldsymbol{\nu}})}\right) \,,
\end{equation}
where $\mathcal{L}$ is the likelihood and $\boldsymbol{\nu}$ denotes all remaining cosmology and nuisance parameters. 
The denominator denotes the likelihood $\mathcal{L}(\hat{\theta},\hat{\boldsymbol{\nu}})$ at the global maximum likelihood estimate (MLE)  or ``best fit'' $(\hat{\theta},\hat{\boldsymbol{\nu}})$. The numerator denotes the likelihood $\mathcal{L}(\theta,\hhat{\boldsymbol{\nu}})$ of the conditional MLE $\hhat{\boldsymbol{\nu}}$ of $\boldsymbol{\nu}$ for a fixed value of $\theta$, which we will occasionally refer to as ``conditional best fit'' for simplicity. In other words, the numerator maximizes the likelihood over all other parameters ${\boldsymbol{\nu}}$ at a fixed value of $\theta$. 

For a Gaussian likelihood $\mathcal{L}$, \cref{eq:PL} indeed follows a $\chi^2$-distribution. Thus the profile likelihood is obtained by ``profiling'' i.e.\ maximizing the likelihood, or equivalently minimizing the $\Delta\chi^2$ over all $\boldsymbol{\nu}$ for a fixed value of $\theta$. The frequentist $68.27\%$\footnote{For brevity, denoted as $68\%$ in the following.} ($95.45\%$) confidence interval is determined using the so-called graphical method, where the interval is defined by the intersection of the profile likelihood with $\Delta\chi^2 = 1$ ($\Delta\chi^2 = 4$, etc.). 
This graphical method guarantees correct frequentist coverage only in the asymptotic limit of a large dataset and far away from any physical boundaries \cite{Neyman:1937uhy, Wilks:1938dza, Wald:1943, Feldman:1997qc}. Since performing a full frequentist Neyman construction, which guarantees correct coverage under all conditions, is prohibitively expensive, we acknowledge that correct frequentist coverage might not be assured in this case. For an exploration of frequentist coverage in a cosmological context see \cite{Herold:2024enb}\footnote{Ref. \cite{Herold:2024enb} found that a $w$CDM model under \Planck\ \texttt{Pliklite} data \cite{Planck:2019nip} seems to violate the asymptotic assumptions. 
However, here we use a more constraining dataset for the $w_0 w_a$CDM model and thus the results of \cite{Herold:2024enb} are not easily transferrable.}.


\section{Results}
\label{sec:results}

In this section, we discuss the frequentist and Bayesian intervals on the CPL parameters $w_0$ and $w_a$ separately, followed by a comparison of both approaches. Our results are summarized in \cref{fig:whisker} and \cref{app:posteriors}. \cref{fig:whisker} shows several dataset combinations, both with and without CMB data. However, our main results here take the CMB as our baseline and focus on constraints when combining this dataset with others. Details on combinations without CMB data are provided in \cref{app:posteriors,app:noCMB} for completeness.

\begin{figure}
    \centering
    \includegraphics[width=\linewidth]{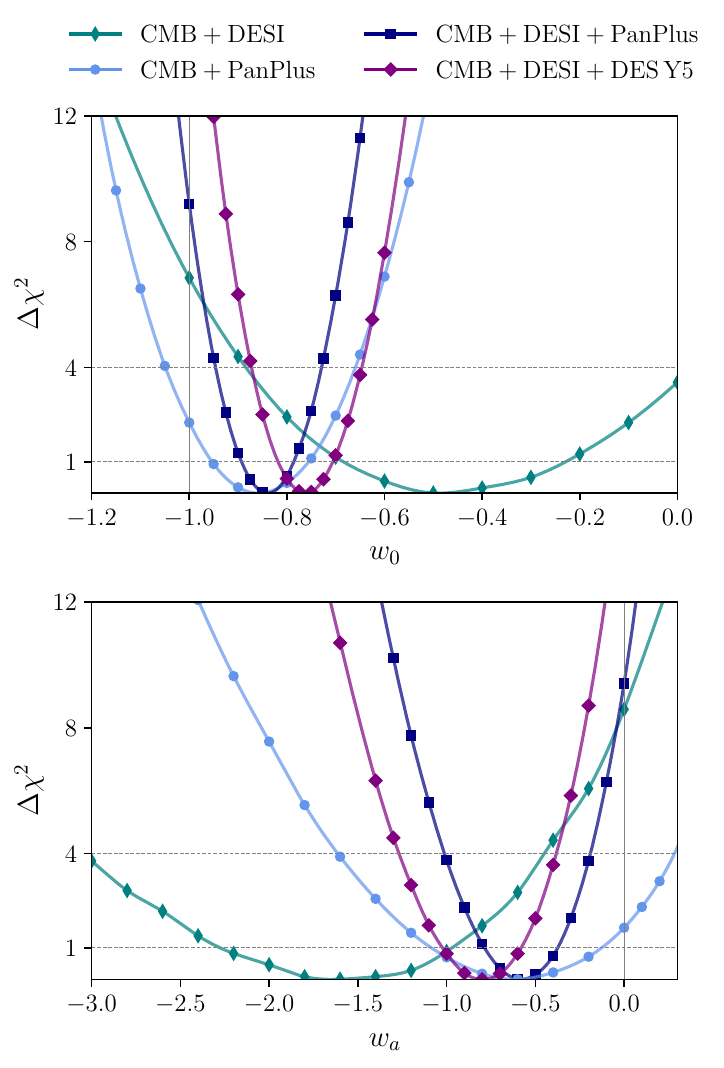}
    \caption{Profile likelihoods for the CPL parameters $w_0$ and $w_a$ (markers) for four different dataset combinations as indicated in the legend. 
    The 68\% confidence intervals are obtained at the intersections of the interpolated curves with $\Delta\chi^2 = 1$. }
    \label{fig:PLs_w0wa}
\end{figure}

The frequentist profile likelihoods of $w_0$ and $w_a$ are shown in \cref{fig:PLs_w0wa} for four different dataset combinations as indicated in the legend. The curves are cubic spline interpolations of the profiles. Then, the $1, \, 2 \sigma$ confidence intervals can be read off from the intersection of the curves with the $\Delta \chi^2 = 1,\, 4$ horizontal dashed lines. Also evident from this figure is the $\chi^2$ improvement over $\Lambda$ provided by $w_0w_a$. $\Lambda$ is marked by the vertical lines at $w_0 = -1,\, w_a = 0$. The $\Delta\chi^2$ at the intersection of the interpolated curves and these vertical lines quantifies the $\Delta\chi^2$ improvement for each data combination when allowing DE parameters to vary under the CPL model. CMB + Pantheon+ sees the smallest improvement, while CMB + DESI + DES Y5 sees the largest. We have checked that these profiles are well fit by a parabola within $1\sigma$ of the minimum $\chi^2$ or the MLE, as expected for a Gaussian likelihood, but show deviations further away from the MLE. Hence correct $95\%$ coverage for the $2\sigma$ constraint may be uncertain. 

In App.~\ref{app:noCMB} we show the profile likelihoods excluding data from the CMB, i.e.\ DESI alone, DESI combined with Pantheon+, and DESI combined with DES Y5.

Our Bayesian MCMC posteriors are shown in \cref{app:posteriors}. Comparing to the Bayesian posteriors of the DESI DR2 results \cite{DESI:2025zgx}, we find small shifts in the mean at the level of $0.2\,\sigma$ and small differences in the size of the credible intervals smaller than $9\%$\footnote{All $\sigma$-differences are computed as $\sqrt{(c_1 - c_2)^2/\sigma^2}$, where $c_1$, $c_2$ are the central values of the two intervals and $\sigma$ is the $68\%$ uncertainty of the Bayesian constraints obtained in this work.}. These small differences with \cite{DESI:2025zgx} might be due to the slightly different priors, different samplers, i.e.\ \texttt{MontePython} (here) vs.\ \texttt{Cobaya} (DESI coll.), different cosmology codes, i.e.\ \texttt{CLASS} (here) vs.\ \texttt{CAMB} \cite{Lewis:1999bs} (DESI coll.), and the different approximations of the neutrino hierarchy, i.e.\ three degenerate masses (here) vs.\ one massive and two massless neutrinos (DESI coll.). 
The only exception is the DESI data alone - in these posteriors (fifth row in \cref{tab:constraints}), we find constraints on $w_0$ that are $0.6\sigma$ higher and constraints on $\Omega_\mathrm{m}$ within CPL that are $0.5\sigma$ higher than in \cite{DESI:2025zgx}; additionally our $1\sigma$ credible intervals are 40\% larger. This difference is most likely caused by the different choice of prior compared to \cite{DESI:2025zgx}, which becomes important for DESI data alone due to the lower constraining power compared to the other dataset combinations.

\begin{figure*}
   \centering
    \includegraphics[width=1\linewidth]{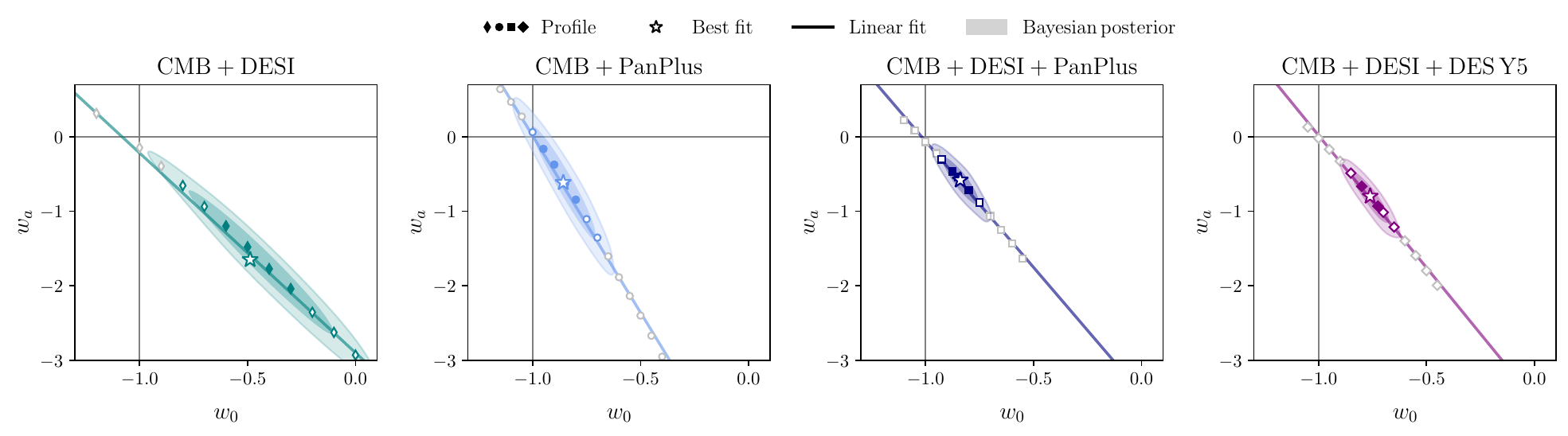}
    \caption{ 
    A comparison of Bayesian posteriors (shaded contours) and profile likelihoods (markers) on the $w_a - w_0$ plane. Markers show the conditional best-fit values $\hhat{w}_a$ of $w_a$ that maximize the likelihood for fixed values of $w_0$ for four different dataset combinations, as indicated in the legend. 
    This is effectively a profile likelihood on $w_0$ projected onto the $w_a-w_0$ plane. 
    Colorful, filled markers show points in the $w_0$ profile likelihood within the $1\sigma$ confidence region, colorful open markers are within $2\sigma$ and gray open markers are beyond this region. These can be compared to the top panel of \cref{fig:PLs_w0wa}, but note that here we plot a subset of the profile points to avoid visual crowding. The extents of these $1\sigma$ and $2\sigma$ profile points closely align with the $1\sigma$ and $2\sigma$ posterior contours, up to the discreteness of the profile points and projection effects of going from 2D posteriors to 1D marginalized limits.
    The lines are linear fits to the profiles and open stars indicate the global best-fit values of $(\hat{w}_0, \hat{w}_a)$. 
    The two approaches are not only centered on the same point, but also trace the same degeneracy directions for all datasets. 
    }
    \label{fig:post_prof_w0wa}
\end{figure*}
Comparing the Bayesian and frequentist intervals obtained in this work, we find excellent agreement on the 1D parameter limits across the two statistical methodologies, as shown in \cref{fig:whisker}. \cref{fig:post_prof_w0wa} overlays the Bayesian posteriors (shaded contours) and the profile likelihoods (filled markers) on the $w_a-w_0$ plane by projecting a profile likelihood on $w_0$ onto this 2D parameter space. Not only are the contours centered at the best fits (marked by open stars), but the linear fits to the profiles on this plane also recover the degeneracy directions of the posteriors showing excellent agreement between the two statistical methods beyond just the 1D limits. 

For $w_0$ and $w_a$, the means of our posteriors agree within $0.2\,\sigma$ with the MLE for all dataset combinations. Moreover, the Bayesian and frequentist error bars agree within $8\%$.\footnote{\cref{app:posteriors} additionally compares $2\sigma$ posteriors and profile-likelihood constraints. We advise caution against taking these at face value as the profiles are no longer quadratic in this extended region, indicating a breakdown of the asymptotic assumption such that a $\Delta \chi^2 = 4$ cutoff no longer guarantees coverage of a $95.45\%$ confidence interval. Regardless, the size of the Bayesian and frequentist $2\sigma$ intervals agree within 10\%.}
We thus find consistent constraints of the parameters $w_0$ and $w_a$ from both approaches. Given the improvement in $\chi^2$ when opening up the $\Lambda$CDM parameter space by varying $w_0$ and $w_a$ (e.g.\ Tab.\ VI in \cite{DESI:2025zgx}, see also \cref{app:chi2contr}), a significant impact of prior-volume effects was not expected. Our results validate this interpretation. Altogether, we find consistent constraints on $w_0$ and $w_a$ from both approaches and that the preference for $w_0 > -1$ and $w_a < 0$ is robust under a change of statistical method. 

Beyond the CPL parameters, we recover the shifts in the MLEs of the \LCDM parameters observed in Bayesian explorations. We confirm the trend in $H_0$ observed in previous Bayesian analyses \cite{Tang:2024lmo,DESI:2024mwx,DESI:2025zgx, RoyChoudhury:2025dhe} --- CPL lowers $H_0$, worsening the Hubble tension \cite{Kamionkowski:2022pkx, Poulin:2023lkg, CosmoVerse:2025txj}, and a late-universe resolution to the Hubble tension remains elusive. For CMB and DESI data, we find $H_0 = 68.5 \pm 0.3\, \mathrm{km}\,\mathrm{s}^{-1} \mathrm{Mpc}^{-1}$ within $\Lambda$CDM compared to $H_0 = 64.1\pm 2.0\, \mathrm{km}\,\mathrm{s}^{-1} \mathrm{Mpc}^{-1}$ within CPL. Including SN data, which prefer lower values of $\Omega_\mathrm{m}$ within CPL, raises $H_0$ to a level similar to \LCDM\ --- for CMB, DESI and Pantheon+ we find $H_0 = 68.3\pm 0.3 \, \mathrm{km}\,\mathrm{s}^{-1} \mathrm{Mpc}^{-1}$ within \LCDM\ and $H_0 = 67.6 \pm 0.6\, \mathrm{km}\,\mathrm{s}^{-1} \mathrm{Mpc}^{-1}$ within CPL. The matter fraction $\Omega_{\rm m}$ shifts to slightly higher values within CPL at consensus for all combinations except CMB and Pantheon+, the combination with the smallest tension on this parameter within \LCDM, as will be discussed further in the next section.

\section{Discussion}
\label{sec:discussion}

\subsection{\texorpdfstring{$\Omega_\mathrm{m}$}{Om} discrepancies across \texorpdfstring{\LCDM}{LCDM} and CPL}
\label{sec:Omm}

\begin{figure*}[h]
    \centering
    \includegraphics[width=0.95\linewidth]{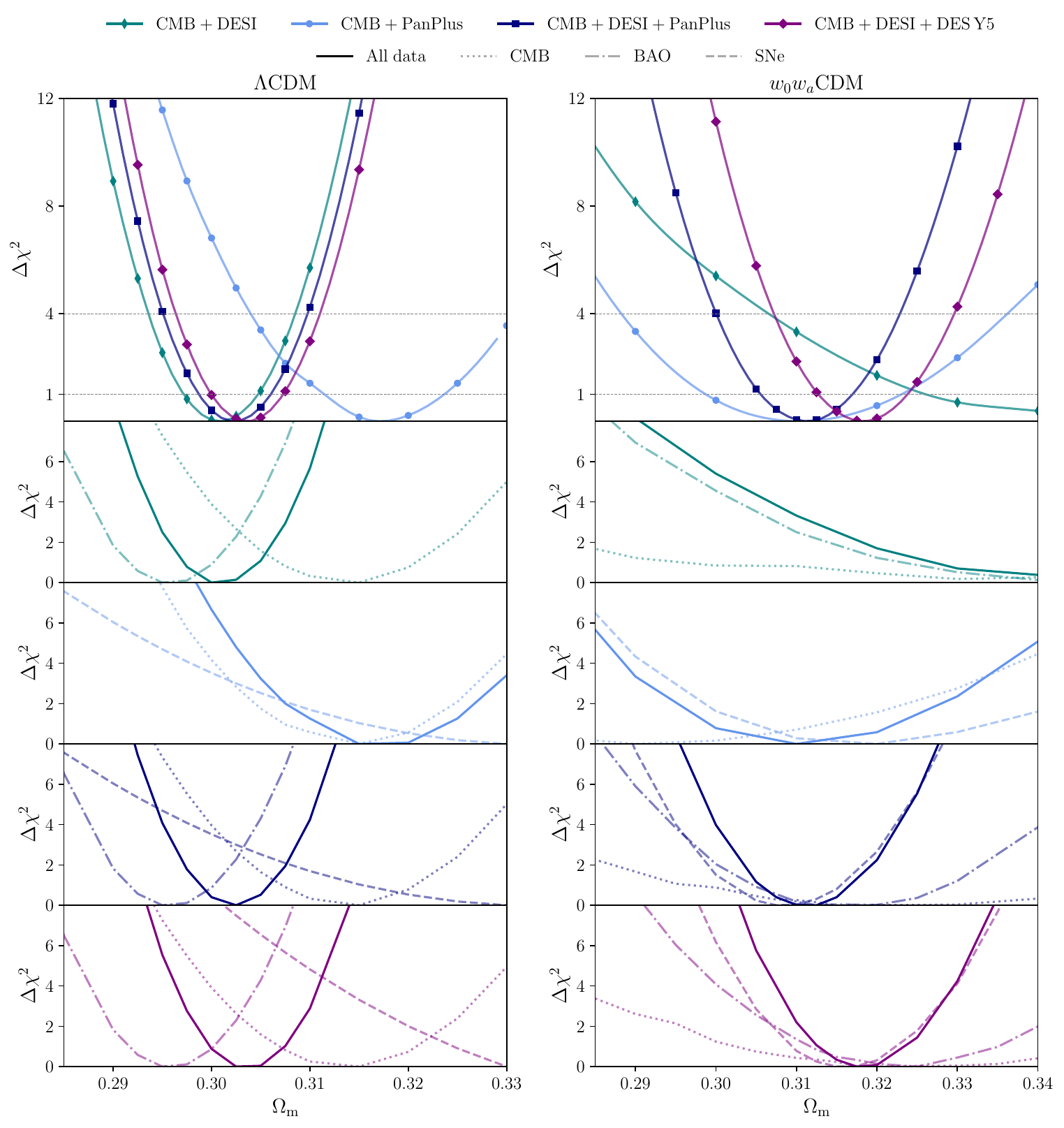}
    \caption{\textit{Top:} Profile likelihoods in $\Omega_\mathrm{m}$ (markers) for $\Lambda$CDM (left) and  $w_0w_a$CDM (right) for the different dataset combinations indicated in the legend. 
    The solid curves interpolate the profile, their intersections with $\Delta\chi^2 = 1,\, 4$ defining the $1,\, 2\sigma$ regions.
    \textit{Other panels:} For each dataset combination, indicated by the colors in the legend, the respective panel shows the total $\Delta\chi^2$ (solid) along with the contributions from the individual datasets: CMB (dotted), DESI BAO (dash-dotted) and SNe (dashed), c.f.\ Eq.~\ref{eq:chi2_contributions}. Each curve is individually adjusted by subtracting out its minimum $\chi^2$. It becomes evident that the individual datasets show a mild discrepancy in their preferred values of $\Omega_\mathrm{m}$ within $\Lambda$CDM (left), which disappears for $w_0w_a$CDM (right).}
    \label{fig:PLs_Om}
\end{figure*}
Recent studies have shown that mild discrepancies on the preferred value of the matter density $\Omega_\mathrm{m}$ within \LCDM contribute to the preference for $w_0$, $w_a$ differing from a cosmological constant \cite{Tang:2024lmo, Berghaus:2024kra, DES:2025bxy}. DESI BAO data prefer smaller values of $\Omega_\mathrm{m} = 0.298 \pm 0.009$ \cite{DESI:2025zgx} than the CMB, $\Omega_\mathrm{m} = 0.315 \pm 0.007$ (for \Planck\ 2018 \cite{Planck:2018vyg}), while SNe prefer greater matter densities $\Omega_\mathrm{m} = 0.352 \pm 0.017$ (for DES Y5 \cite{DES:2024jxu}). 
BAO and SNe both probe cosmic distances with an overlapping range in redshift. As flat \LCDM can only tune $\Omega_{\rm m}$ to fit the late-universe expansion rate, this model does not contain the flexibility to simultaneously fit both datasets. 

Here, we explore this divergence on $\Omega_{\rm m}$ constraints with profile likelihoods. Profile likelihoods of $\Omega_\mathrm{m}$ within the $w_0 w_a$CDM and $\Lambda$CDM models are shown in the top panels of \cref{fig:PLs_Om} for different dataset combinations. Our credible and confidence intervals are shown in \cref{fig:whisker} and in \cref{app:posteriors}. Again, we find excellent agreement between the two approaches. The means and MLEs agree within $0.1\sigma$ and the Bayesian and frequentist error bars agree within $9\%$.

In order to explore which datasets drive the constraints, we consider the contributions to the $\chi^2$ from the individual datasets. The total $\chi^2$ can be written as a sum of the individual contributions 
\begin{equation}
	\label{eq:chi2_contributions}
	\chi^2_\mathrm{tot} = \chi^2_\mathrm{CMB} + \chi^2_\mathrm{BAO} + \chi^2_\mathrm{SNe} \,,
\end{equation}
consisting of CMB (\Planck\ primary and combined \Planck\ and ACT lensing), BAO (DESI) and SNe (either Pantheon+ or DES Y5). The individual contributions to the total $\chi^2$ are shown in the four lower panels of \cref{fig:PLs_Om}. Note that the $\Delta\chi^2$ of the individual datasets (CMB, BAO, SNe) are obtained by fixing $\Omega_{\rm m}$ and maximizing the likelihood for the combination of datasets over all other parameters. The profile likelihood is then fragmented into its contributions from the individual datasets. This breakdown of $\chi^2$ contributions is not obtained from fits to the individual datasets, so shifts are expected across rows. 

Within $\Lambda$CDM (left panels), we recover the Bayesian preference for different values of $\Omega_\mathrm{m}$ from individual datasets. The second panel (green lines) illustrates how the frequentist confidence interval $\Omega_\mathrm{m} = 0.3011 \pm 0.0038$ (solid line) under CMB and DESI data results from a compromise between the lower value preferred by DESI, $\Omega_\mathrm{m} \sim 0.295$ (dash-dotted), and the higher value preferred by the CMB,  $\Omega_\mathrm{m} \sim 0.315$ (dotted). In the third panel (light-blue lines), we show the contributions to the $\Delta\chi^2$ from CMB and Pantheon+ SNe. The value of $\Omega_\mathrm{m} \sim 0.33$ preferred by Pantheon+ (dashed line) lies only marginally higher than the CMB-preferred value. The fifth and sixth panels show $\Delta\chi^2$-data of CMB, DESI combined with SN data from Pantheon+ (dark blue) or DES Y5 (purple), respectively. In both cases, the $\Omega_\mathrm{m}$ values preferred by DESI are slightly lower than the ones preferred by CMB and SNe, and the total $\chi^2$ thus finds a compromise between the individual contributions. 

This compromise from forcing a combined fit under \LCDM comes at a cost to $\chi^2$ for the individual datasets.  In \cref{fig:PLs_Om}, reading off the $\Delta \chi^2$ for individual datasets at the minimum of the solid curve gives the cost to $\chi^2$ for that dataset while forcing a common cosmology. This provides a lower bound to the total cost to $\chi^2$ for each individual dataset, which should be compared to an individual fit to that data (allowing for individual cosmologies) as opposed to a combined fit. For example, from the bottom panel, the fit to DES Y5 is worsened by $\Delta \chi^2 \sim 7$, for DESI by $\Delta \chi^2 \sim 3$ and for CMB by $\Delta \chi^2 \sim 2$.

This mild discrepancy on the preferred $\Omega_\mathrm{m}$ values disappears within $w_0w_a$CDM (right panels). The second panel (green lines) illustrates how the individual contributions to the constraint $\Omega_\mathrm{m} = 0.353 \pm 0.021$ (solid line) under CMB and DESI data now agree on the preferred value of $\Omega_\mathrm{m}$, although the constraining power is degraded by the introduction of the two additional parameters $w_0$ and $w_a$. We find similar features for the third through fifth panels: the individual contributions agree on the central values of $\Omega_\mathrm{m} \sim 0.32$. Again, note that the contributions of the individual datasets (e.g.\ CMB, dashed lines) in the different panels can differ due to the breaking of parameter degeneracies facilitated by the remaining datasets.

This illustrates how opening up the parameter space by varying $w_0$, $w_a$ allows for a region in parameter space that simultaneously optimizes the fit for all datasets, accompanied by an improvement in $\chi^2$ along with an increase in the uncertainty. This is in agreement with previous work using Bayesian approaches \cite{Tang:2024lmo,Berghaus:2024kra, DES:2025bxy}.

We also show the contributions of the individual datasets to the total $\chi^2$ for profile likelihoods on $w_0$ and $w_a$ in \cref{app:chi2contr}. As expected, the constraints on both $w_0$ and $w_a$ are dominated by the SNe and BAO data, while CMB data play a subdominant role. Nevertheless, \Planck\ primary CMB and \Planck\ and ACT lensing show a mild preference of $w_0 > -1$ and $w_a < 0$, as was previously noted in \cite{Planck:2018vyg}. We show a breakdown of the improvement in $\chi^2$ in \cref{app:chi2contr} for all dataset combinations. Evidently, all datasets considered in this work show a mild preference for the $w_0w_a$ model.

\begin{figure*}
   \centering
    \includegraphics[width=\linewidth]{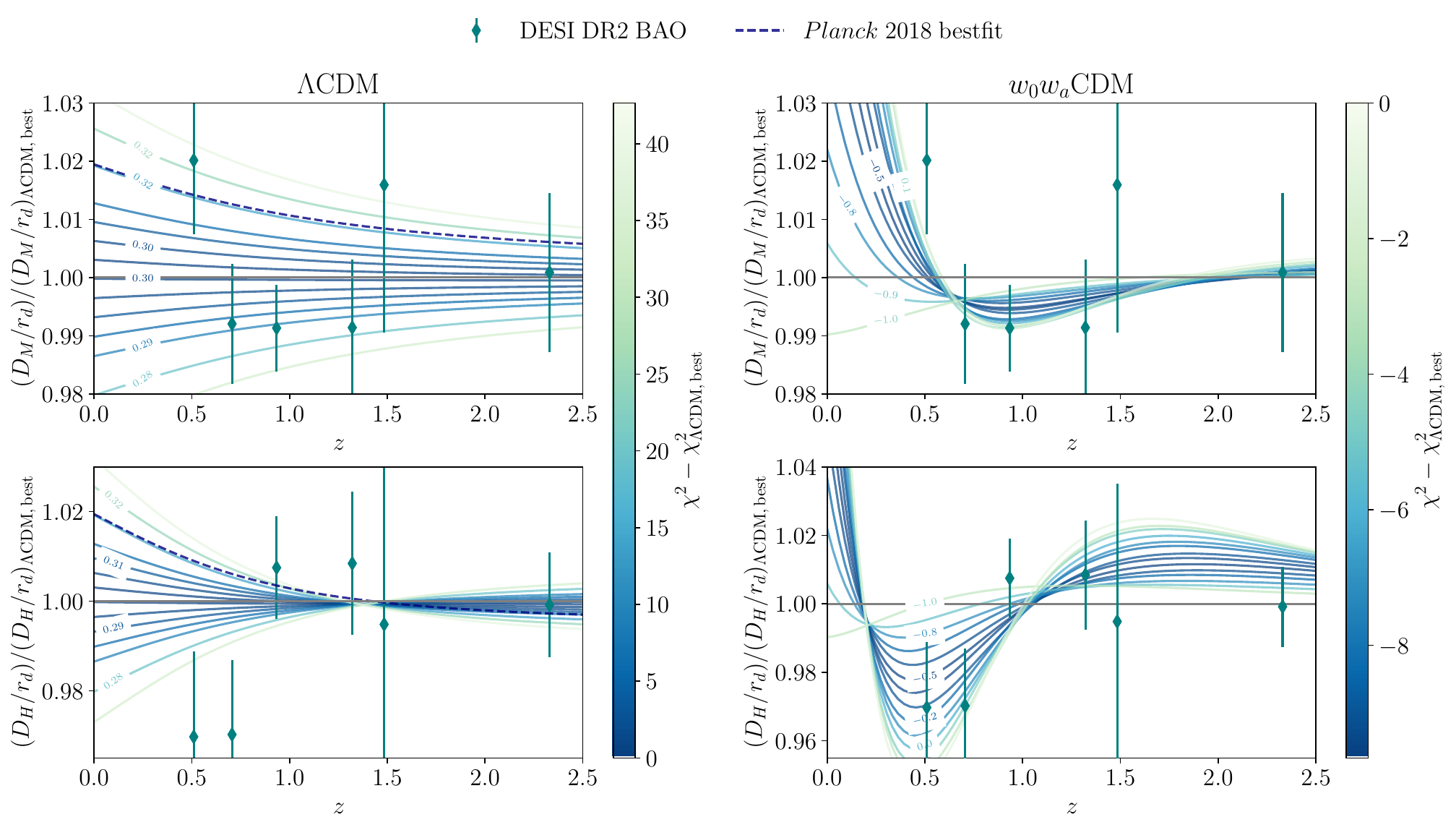}
    \caption{
    The curves show $\Lambda$CDM conditional best-fit cosmologies for fixed values of $\Omega_\mathrm{m} \in [0.28, 0.325]$ (left) and $w_0w_a$CDM conditional best-fit cosmologies for fixed values of $w_0 \in [-1, 0.3]$ (right).
    The numbers along the curves denote the values of $\Omega_{\rm m}$ (left) and $w_0$ (right) for reference, with text color matching the color of the corresponding curve. 
    The transverse $D_M(z)/r_d$ (top row), and line-of-sight $D_H(z)/r_d$ (bottom row) distances are shown relative to the $\Lambda$CDM global best-fit cosmology for CMB and DESI data.
    For comparison, the markers denote the DESI DR2 BAO measurements \cite{DESI:2025zgx}. 
    The colorbar indicates the difference in goodness of fit w.r.t.\ the $\Lambda$CDM best-fit cosmology $\chi^2-\chi^2_{\Lambda\mathrm{CDM,\, best}}$. 
    The \Planck\ 2018 best-fit cosmology \cite{Planck:2018vyg} (blue dashed) prefers higher values of $\Omega_\mathrm{m}$ than DESI within $\Lambda$CDM.
    This illustrates how varying $\Omega_\mathrm{m}$ can only modulate the slope and amplitude of $D_M(z)/r_d$ and $D_H(z)/r_d$, while the $w_0w_a$ model introduces more freedom to fit the DESI BAO data. }
    \label{fig:spaghetti}
\end{figure*}

We further explore the impact of different $\Omega_\mathrm{m}$ values on BAO data in \cref{fig:spaghetti}. We show the two cosmological distances constrained by BAO data: 
\begin{equation}
    \frac{D_M}{r_\d} = \frac{c}{r_\d} \int_0^z \frac{\d z}{H(z)} \,,
    \qquad \frac{D_H}{r_\d} = \frac{1}{r_\d\, H(z)} \,,
\end{equation}
the transverse (angular diameter) distance $D_M(z)$ and the line-of-sight distance $D_H(z)$, where $H(z) \approx H_0 \sqrt{\Omega_\mathrm{m}(1+z)^3 + \Omega_\mathrm{DE}(z)}$ is the expansion rate in the late universe. BAO data are sensitive to these quantities relative to the sound-horizon size $r_\d$ at the drag epoch. 

The left panels of \cref{fig:spaghetti} show the conditional best-fit cosmologies under CMB and DESI data for different fixed values of $\Omega_\mathrm{m} \in [0.28, 0.325]$, i.e.\ we minimize the $\chi^2$ for each fixed $\Omega_\mathrm{m}$ over all remaining cosmology and nuisance parameters. We color-code the lines according to the $\chi^2$ under CMB and DESI data relative to the global MLE $\chi^2_{\Lambda\mathrm{CMB\, best}}$ under the same data (horizontal gray line). 
This demonstrates how varying $\Omega_\mathrm{m}$ can only modulate the slope and amplitude of $D_M(z)$ and $D_H(z)$ in order to fit DESI BAO data (markers \cite{DESI:2025zgx}). Moreover, we show the \Planck{} 2018 best-fit cosmology (first column in Table 1 of \cite{Planck:2018vyg}) for comparison (blue dashed line), which prefers higher values of $\hat{\Omega}_\mathrm{m} = 0.316$. This illustrates how DESI data push $\Omega_\mathrm{m}$ to smaller values $\hat{\Omega}_\mathrm{m} = 0.301$ than preferred by the CMB. 

The right panels of \cref{fig:spaghetti} show the conditional best-fit cosmologies under CMB and DESI data for different fixed values of $w_0 \in [-1, 0.3]$, minimized over all remaining cosmology and nuisance parameters. The color of the line indicates the $\chi^2$ relative to the best-fit $\Lambda$CDM cosmology for CMB and DESI data (corresponding to the horizontal gray line). Opening up the parameter space by varying $w_0$ and $w_a$ provides the necessary freedom to fit the DESI BAO data, improving the fit by $\Delta\chi^2 \approx -10$ compared to the best-fit $\Lambda$CDM cosmology.

\subsection{Equation of state at the pivot redshift}
\label{sec:z_pivot}

Projecting the dynamical-DE constraints onto an alternate parameter space defined by a pivot redshift $z_\mathrm{p}$ and equation of state $w_\mathrm{p}$ at this redshift, constraints seem to favor $w_\mathrm{p} \sim -1$ \cite{DESI:2024mwx}.
This raised questions about the robustness of evidence for dynamical DE \cite{Cortes:2024lgw, Efstathiou:2025tie}. 
Here, we explore projections into the $w_\mathrm{p}-z_\mathrm{p}$ parameterization with profile likelihoods. 
\begin{figure}[h]
   \centering
    \includegraphics[width=0.95\linewidth]{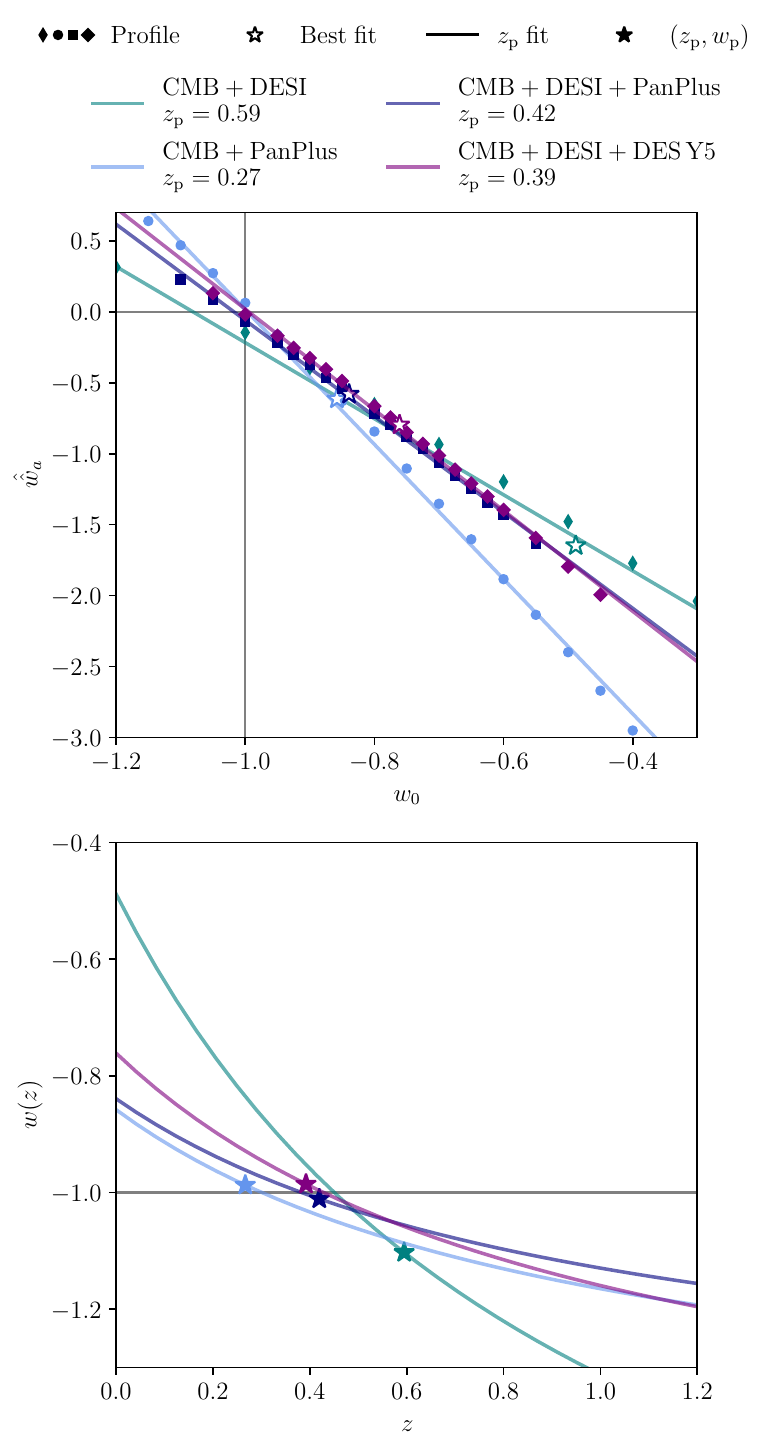}
    \caption{\textit{Top:} Profile likelihoods for a fixed value of $w_0$ with corresponding conditional best fit $\hhat{w}_a$ (markers, same as \cref{fig:post_prof_w0wa}). The open stars indicate the global best-fit values of $(\hat{w}_0, \hat{w}_a)$. Slopes of linear fits to the data are used to obtain the pivot redshifts $z_\mathrm{p}$ using \cref{eq:CPL_reparam}, specified for each dataset in the legend and in \cref{tab:pivot_zs}.  
    \textit{Bottom:} Curves show the best-fit equations of state $w(z)$ as a function of redshift for the datasets indicated in the legend. Stars mark the equations of state $w_\mathrm{p}$ at the pivot redshifts $z_\mathrm{p}$. 
    For all datasets, we find $w_\mathrm{p} \approx -1$. 
    }
    \label{fig:wa_of_w0}
\end{figure}

\cref{eq:CPL} represents an expansion of the equation of state of DE $w(a)$ around the current scale factor $a=1$. Since BAO and SNe data are not particularly sensitive to $a=1$, but to slightly higher redshifts $z\gtrsim 0.1$ ($a\lesssim 0.9$), $w_0$ and $w_a$ are usually strongly correlated. This is illustrated in the top panel of \cref{fig:wa_of_w0}, which shows the conditional MLE $\hhat{w}_a$ of $w_a$ for a fixed value of $w_0$ (colorful markers) for different dataset combinations as indicated in the legend. The MLE for the respective dataset is indicated by an open star. 

As suggested by \cite{Huterer:2000mj} (see also \cite{Albrecht:2006um, Linder:2006xb, Martin:2006vv}), it is therefore useful to define a pivot scale $a_\mathrm{p}$ such that $w_\mathrm{p}$ and $w_a$ decorrelate for the dataset at hand. 
The equation of state $w_\mathrm{p}$ at the pivot scale $a_\mathrm{p}$ can then be expressed as 
\begin{equation}
	\label{eq:CPL_reparam}
	w_\mathrm{p} = w_0 + w_a\, (1-a_\mathrm{p}) \,,
\end{equation}
i.e. the model is reparameterization invariant under a change of pivot scale. Previous studies have found that the equation of state at the pivot redshift under CMB, DESI BAO (and SN) data is consistent with $w_\mathrm{p} = -1$ \cite{DESI:2024mwx, Cortes:2024lgw, DESI:2025zgx}. 

To explore this with both Bayesian and frequentist methods, we compute the pivot scale as detailed in \cite{Albrecht:2006um}. The pivot scale  $a_\mathrm{p}$ is determined as the scale factor that minimizes the standard deviation $\sigma_{w_\mathrm{p}}$ of $w_\mathrm{p}$ in \cref{eq:CPL_reparam}, retrieving the best-constrained direction on the $w_0-w_a$ plane. This yields \cite{Albrecht:2006um}
\begin{equation}
    \label{eq:pivot_scale}
    1-a_\mathrm{p} = - \frac{\mathrm{cov}(w_0 w_a)}{\sigma_{w_a}^2} \,,
\end{equation}
where $\sigma_{w_a}^2$ is the standard deviation of $w_a$ and $\mathrm{cov}(w_0 w_a)$ denotes the covariance between $w_0$ and $w_a$, i.e.\ the off-diagonal elements of the covariance matrix between $w_0$ and $w_a$. We compute the pivot redshifts $1+z_\mathrm{p} = 1/a_\mathrm{p}$ from the covariance matrices obtained from the MCMC runs for the four dataset combinations including the CMB and quote these in the first column of \cref{tab:pivot_zs}. The pivot redshifts are in broad agreement with the ones quoted in \cite{DESI:2025zgx}, which find $z_\mathrm{p} = 0.53$ for CMB and DESI, and $z_\mathrm{p} = 0.31$ for CMB, DESI and DES Y5. 
\begin{table}
\begin{ruledtabular}
\begin{tabular}{l|cc}
Dataset            & $(z_\mathrm{p}, w_\mathrm{p})_\mathrm{Bayes.}$ & $(z_\mathrm{p}, w_\mathrm{p})_\mathrm{freq.}$\\
\hline
CMB, DESI           & $(0.51, -1.00 \pm 0.05)$     & $(0.65, -1.09 \pm 0.10)$ \\
CMB, PanPlus        & $(0.23, -0.97 \pm 0.03)$     & $(0.27, -0.98 \pm 0.03)$ \\
CMB, DESI, PanPlus\ & $(0.31, -0.98 \pm 0.03)$     & $(0.42, -1.01 \pm 0.03)$ \\
CMB, DESI, DES Y5   & $(0.30, -0.95 \pm 0.03)$     & $(0.39, -0.98 \pm 0.03)$ \\
\end{tabular}
\end{ruledtabular}
\\
\caption{The DE equations of state $w_\mathrm{p}$ at the pivot redshifts $z_\mathrm{p}$ obtained from the MCMC covariance matrices through \cref{eq:pivot_scale} (Bayes.) and through decorrelating the conditional MLE $(w_0, \hhat{w}_a)$ through \cref{eq:freq_zp} (freq.).}
\label{tab:pivot_zs}
\end{table}

Analogously, we obtain the pivot redshifts from the frequentist data by finding $z_\mathrm{p}$ such that $\hhat{w}_a$ and $w_\mathrm{p}$ decorrelate. We use a linear fit to determine the inclination $A$ of the conditional MLE $\hhat{w}_a$ as a function of the fixed value of $w_0$ from the top panel of \cref{fig:wa_of_w0}. Comparing to \cref{eq:CPL_reparam} yields the pivot redshift:
\begin{align}
    &\hhat{w}_a = A\, w_0 + \mathrm{const.} \,,\nonumber \\
    &\Rightarrow z_\mathrm{p} = -(1+A)^{-1} \,.
    \label{eq:freq_zp}
\end{align}
These linear fits are shown by the colored lines in the top panel of \cref{fig:wa_of_w0} and the frequentist $z_\mathrm{p}$ are quoted in the right column of \cref{tab:pivot_zs}. The values of the pivot redshifts obtained with the conditional MLE $\hhat{w}_a$ are in good agreement with the ones obtained from the posterior covariance matrices. The differences in $z_\mathrm{p}$ obtained from Bayesian and frequentist data are not surprising since the determination of $z_\mathrm{p}$ includes a large uncertainty, especially when the posterior contours do not show a strong degeneracy. 

Using $z_\mathrm{p}$ and the MLEs $(\hat{w_0}, \hat{w_a})$, we compute the equation of state $w_\mathrm{p}$ at the pivot redshift via \cref{eq:CPL_reparam}. We show $(z_\mathrm{p}, w_\mathrm{p})$ as the stars in the bottom panel of \cref{fig:wa_of_w0}, along with the best-fit DE equation of state $w(z)$. The numerical values are quoted in \cref{tab:pivot_zs}. The uncertainties are obtained from MCMC chains or profile likelihoods in $w_\mathrm{p}$ with $z_\mathrm{p}$ fixed to the value inferred from the respective dataset and statistical method.

As noted in previous studies \cite{DESI:2024mwx, Cortes:2024lgw, DESI:2025zgx, Efstathiou:2025tie} based on Bayesian methods, we find that the equations of state at the pivot redshifts are close to $w_\mathrm{p} \approx -1$. Thus current data seem to be more sensitive to the derivative, $w_a \neq 0$, rather than a mean offset $w_\mathrm{p} \neq -1$. Hence preferred models start off in the ``phantom regime'', $w(z)<-1$, and transition to the non-phantom regime around $z_\mathrm{p}$, which can be interpreted as a new ``coincidence problem'' \cite{Cortes:2024lgw}. 

We emphasize that finding $w_{\rm p} \simeq -1$ does not imply consistency with a cosmological constant. The frequentist $w_{\rm p}$ constraints in \cref{tab:pivot_zs} are simply a reparameterization of $w_0w_a$ constraints and are distinct from a cosmological constant, as shown by the $(w_0\,, w_a)$ stars in the top panel of \cref{fig:wa_of_w0}, which correspond to the $(z_{\rm p}\,, w_{\rm p})$ stars in the bottom panel. Specifically, note that the location of the best fit on the $w_0-w_a$ plane is not at $\Lambda$. Indeed, the best-fit $w_{\rm p}$ carries its resultant $\chi^2$ improvement over \LCDM, as reported in \cref{tab:chi2}, and incorporates a dynamical DE which should be read as the full $w(z)$ curve, not just its value $w_{\rm p}$ at $z_{\rm p}$. 


\section{Conclusions}
\label{sec:conclusions}

In this paper, we complemented the Bayesian credible intervals on the CPL parameters $w_0$ and $w_a$ with frequentist confidence intervals based on the graphical profile-likelihood method. We found that frequentist constraints are in excellent agreement with Bayesian constraints (Fig.~\ref{fig:whisker}). Shifts of the best fits compared to the posterior means are smaller than $0.2\,\sigma$ and the sizes of the frequentist and Bayesian intervals agree within $9\%$. This reaffirms the robustness of the constraints on dynamical DE under a change of statistical method and shows no impact from prior-volume effects. 

We further assessed which datasets drive the constraints by exploring the contributions to the $\chi^2$ from the individual datasets. While all datasets show a mild but consistent preference for $w_0 > -1$ and $w_a < 0$, the constraining power is dominated by the BAO and SN data. The improvement in fit compared to \LCDM\ is strongest in DES Y5 data ($\Delta \chi^2 \simeq -11$) when combined with CMB and BAO  for 2 extra degrees of freedom (see \cite{Efstathiou:2024xcq,DES:2025tir} for a discussion of this SNe sample). DESI, CMB and Pantheon+ individually share a similar level of preference, $-5 < \Delta \chi^2 < -3$. This preference for the $w_0w_a$ model is driven by mild discrepancies in the preferred values of the matter fraction $\Omega_\mathrm{m}$ of the different datasets (\cref{fig:PLs_Om}), which are resolved within $w_0w_a$, in agreement with previous Bayesian works \cite{Berghaus:2024kra, Tang:2024lmo}. Varying $\Omega_\mathrm{m}$ within \LCDM\ can only modulate the amplitude and slope of the BAO observables $D_M$, $D_A$, while the $w_0w_a$ model allows for more freedom to fit the DESI BAO data (\cref{fig:spaghetti}).

We also confirmed that the equations of state $w_\mathrm{p}$ of DE at the pivot redshifts, which correspond to the best-constrained $w(z)$ for the respective data, are close to $-1$ (\cref{fig:wa_of_w0}) with a frequentist approach. Thus, the data appear to be more sensitive to the derivative of $w(z)$ than a mean offset from $w=-1$, also in agreement with Bayesian investigations \cite{DESI:2024mwx, Cortes:2024lgw, DESI:2025zgx, Efstathiou:2025tie}. 

Altogether, in this paper, we employed primarily frequentist methods and found that our results align remarkably well with those reported in the literature using predominantly Bayesian approaches \cite{DESI:2025zgx}. Future data will be crucial in determining whether this challenge to the $\Lambda$CDM model persists.


\vspace{3mm}
\textit{Acknowledgements:} 
We are grateful for comments from Marc Kamionkowski that substantially improved this work. 
We are also grateful for helpful discussions with Graeme Addison, Dillon Brout, Elisa Ferreira, Eric Linder, Jessica Muir, Shivam Pandey, Hiranya Peiris, Adam Riess and Zach Weiner. 
LH was supported by a William H. Miller fellowship.  
TK was supported by the Kavli Institute for Cosmological Physics at the University of
Chicago through an endowment from the Kavli Foundation. 
This work was carried out at the Advanced Research Computing at Hopkins (ARCH) core facility (\texttt{arch.jhu.edu}), which is supported by the National Science Foundation (NSF) grant number OAC1920103. 
Computing resources were also provided by the University of Chicago Research Computing Center through the Kavli Institute for Cosmological Physics. 
LH is grateful for useful discussions during a MIAPbP workshop, which is funded by the Deutsche Forschungsgemeinschaft (DFG) under Germany's Excellence Strategy – EXC-2094 – 390783311.


\appendix

\vspace{5mm}

\section{Bayesian posteriors and frequentist constraints}
\label{app:posteriors}

\begin{figure*}
   \centering
    \includegraphics[width=1\linewidth]{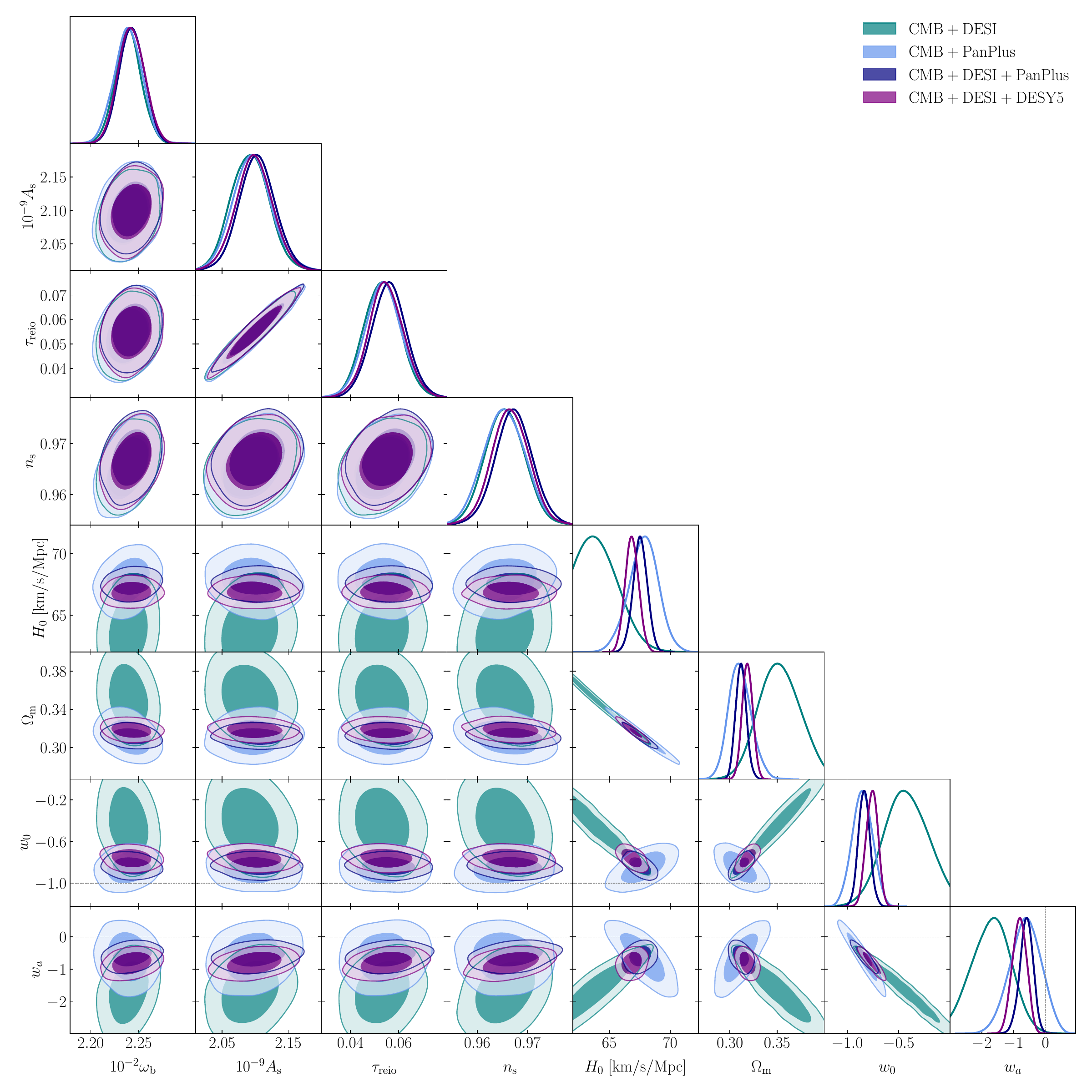}
    \caption{Posterior corner plot of the cosmological parameters within the $w_0w_a$CDM model for four different dataset combinations as indicated in the legend. CMB denotes \Planck\ primary CMB combined with \Planck+ACT lensing.
    }
    \label{fig:posterior_w0wa}
\end{figure*}

\cref{fig:posterior_w0wa} shows the posterior corner plot of the cosmological parameters for the $w_0 w_a$CDM model for various dataset combinations. Our constraints are in good agreement with the ones from the DESI collaboration \cite{DESI:2025zgx}. \cref{tab:constraints} compares these $68\%$ Bayesian to the frequentist constraints as well as the DESI DR2 constraints where available. In \cref{tab:constraints_2sigma}, we additionally quote the $95\%$ constraints.

\begin{table*}
\begin{ruledtabular}
\begin{tabular}{cl|ccc|c}
Data & Statistic & $w_0$                     & $w_a$                     & $\Omega_\mathrm{m}$ (CPL)    & $\Omega_\mathrm{m}$ ($\Lambda$CDM)\\
\hline
\multirow{3}{*}{CMB, DESI\ } 
& Bayesian~\cite{DESI:2025zgx}   & $-0.42 \pm 0.21$          & $-1.75 \pm 0.58$          & $0.353 \pm 0.021$            & $0.303 \pm 0.004$ \\
& Bayesian                       & $-0.42 \pm 0.22$          & $-1.72_{-0.59}^{+0.67}$   & $0.353 \pm 0.022$            & $0.301 \pm 0.004$ \\
& Frequentist                    & $-0.49_{-0.20}^{+0.26}$   & $-1.65_{-0.63}^{+0.68}$   & $0.351_{-0.025}^{+0.021}$    & $0.301 \pm 0.004$\\
\hline
\multirow{2}{*}{\shortstack{CMB,\\ Pantheon+}} 
& Bayesian                       & $-0.85 \pm 0.10$        & $-0.63 \pm 0.49$          & $0.311_{-0.013}^{+0.011}$    & $0.317_{-0.007}^{+0.006}$ \\
& Frequentist                    & $-0.86 \pm 0.10$        & $-0.61_{-0.47}^{+0.49}$   & $0.310_{-0.012}^{+0.013}$    & $0.317 \pm 0.006$\\
\hline
\multirow{3}{*}{\shortstack{CMB, DESI,\ \\ Pantheon+}} 
& Bayesian~\cite{DESI:2025zgx}  & $-0.838 \pm 0.055$        & $-0.62^{+0.22}_{-0.19}$   & $0.311 \pm 0.006$          & $-$ \\
& Bayesian                      & $-0.835 \pm 0.055$        & $-0.60_{-0.19}^{+0.22}$   & $0.312 \pm 0.006$          & $0.302 \pm 0.004$ \\
& Frequentist                   & $-0.839 \pm 0.054$        & $-0.58 \pm 0.21$          & $0.311 \pm 0.006$          & $0.302 \pm 0.004$\\
\hline
\multirow{3}{*}{\shortstack{CMB, DESI,\\ DES Y5}}
& Bayesian~\cite{DESI:2025zgx}  & $-0.752\pm 0.057$         & $-0.86^{+0.23}_{-0.20}$   & $0.319 \pm 0.006$          & $-$ \\
& Bayesian                      & $-0.758 \pm 0.057$        & $-0.82 \pm 0.22$          & $0.319 \pm 0.006$          & $0.304 \pm 0.004$ \\
& Frequentist                   & $-0.761_{-0.057}^{+0.055}$ & $-0.80_{-0.23}^{+0.22}$ & $0.318 \pm 0.006$             & $0.304 \pm 0.004$\\
\hline
\multirow{3}{*}{DESI\ } 
& Bayesian~\cite{DESI:2025zgx}   & $-0.48_{-0.17}^{+0.35}$   & $<-1.34$             & $0.352_{-0.018}^{+0.041}$            & $0.298 \pm 0.009$ \\
& Bayesian                       & $-0.21_{-0.42}^{+0.43}$   & $-2.6 \pm 1.4$       & $0.382_{-0.042}^{+0.050}$            & $0.298 \pm 0.009$ \\
& Frequentist                    & $-0.19_{-0.41}^{+0.45}$   & $-2.7_{-1.5}^{+1.4}$ & $0.388_{-0.047}^{+0.045}$            & $0.297 \pm 0.009$\\
\hline
\multirow{3}{*}{\shortstack{DESI,\\ Pantheon+}}
& Bayesian~\cite{DESI:2025zgx}   & $-0.888_{-0.064}^{+0.055}$  & $-0.17 \pm 0.46$          & $0.298_{-0.011}^{+0.025}$   & $-$ \\
& Bayesian                       & $-0.878_{-0.067}^{+0.058}$  & $-0.27 \pm 0.44$          & $0.304_{-0.012}^{+0.021}$   & $0.305 \pm 0.008$ \\
& Frequentist                    & $-0.882_{-0.060}^{+0.065}$ & $-0.25_{-0.44}^{+0.43}$ & $0.307_{-0.019}^{+0.014}$ & $0.305 \pm 0.008$\\
\hline
\multirow{3}{*}{\shortstack{DESI,\\ DES Y5}}
& Bayesian~\cite{DESI:2025zgx}   & $-0.781_{-0.076}^{+0.067}$          & $-0.72 \pm 0.47$          & $0.319_{-0.011}^{+0.017}$            & $-$ \\
& Bayesian                       & $-0.779_{-0.076}^{+0.068}$          & $-0.75 \pm 0.45$          & $0.321_{-0.012}^{+0.017}$            & $0.311 \pm 0.008$ \\
& Frequentist                    & $-0.782_{-0.069}^{+0.072}$          & $-0.73 \pm 0.45$          & $0.321_{-0.015}^{+0.013}$            & $0.310 \pm 0.008$\\

\end{tabular}
\end{ruledtabular}
\caption{\textbf{68\%} Bayesian credible intervals (mean $\pm 1\sigma$) and frequentist confidence intervals (MLE $\pm 1\sigma$) for $w_0$, $w_a$ {\rm and } $\Omega_\mathrm{m}$ within the CPL model, and $\Omega_\mathrm{m}$ within $\Lambda$CDM. We find excellent agreement between the Bayesian and frequentist intervals. For comparison, we show the constraints published within the DESI DR2 results \cite{DESI:2025zgx} if available.
}
\label{tab:constraints}
\end{table*}


\begin{table*}
\begin{ruledtabular}
\begin{tabular}{cl|ccc|c}
Data & Statistic & $w_0$                     & $w_a$                     & $\Omega_\mathrm{m}$ (CPL)    & $\Omega_\mathrm{m}$ ($\Lambda$CDM)\\
\hline
\multirow{2}{*}{CMB, DESI\ } 
& Bayesian                       & $-0.42_{-0.42}^{+0.43}$ & $-1.72_{-1.29}^{+1.18}$ & $0.353_{-0.041}^{+0.043}$ & $0.301_{-0.007}^{+0.008}$\\
& Frequentist                    & $-0.49_{-0.40}^{+0.52}$ & $-1.65_{-1.40}^{+1.20}$ & $0.351_{-0.044}^{+0.046}$ & $0.301_{-0.007}^{+0.008}$\\
\hline
\multirow{2}{*}{\shortstack{CMB,\\ Pantheon+}}
& Bayesian                      & $-0.85_{-0.19}^{+0.19}$ & $-0.63_{-0.97}^{+0.93}$ & $0.311_{-0.023}^{+0.025}$ & $0.317_{-0.012}^{+0.013}$\\
& Frequentist                   & $-0.86_{-0.19}^{+0.20}$ & $-0.61_{-1.00}^{+0.90}$ & $0.310_{-0.022}^{+0.026}$ & $0.317_{-0.013}^{+0.014}$\\
\hline
\multirow{2}{*}{\shortstack{CMB, DESI,\ \\ Pantheon+}}
& Bayesian                      & $-0.835_{-0.108}^{+0.115}$ & $-0.60_{-0.43}^{+0.39}$ & $0.312_{-0.011}^{+0.012}$ & $0.303 \pm 0.008$\\
& Frequentist                   & $-0.839_{-0.107}^{+0.110}$ & $-0.58_{-0.43}^{+0.39}$ & $0.311 \pm 0.011$         & $0.302 \pm 0.007$\\
\hline
\multirow{2}{*}{\shortstack{CMB, DESI,\\ DES Y5}}
& Bayesian                      & $-0.758_{-0.110}^{+0.109}$ & $-0.82_{-0.44}^{+0.42}$ & $0.319 \pm 0.011$ & $0.304 \pm 0.007$ \\
& Frequentist                   & $-0.761_{-0.111}^{+0.115}$ & $-0.80_{-0.47}^{+0.42}$ & $0.318 \pm 0.011$ & $0.304 \pm 0.007$\\
\hline
\multirow{2}{*}{DESI\ } 
& Bayesian                    & $-0.21 \pm 0.76$        & $-2.6_{-2.6}^{+2.9}$ & $0.381_{-0.098}^{+0.082}$ & $0.298 \pm 0.017$ \\
& Frequentist                 & $-0.19_{-0.80}^{+0.90}$ & $-2.7_{-3.1}^{+2.9}$ & $0.388_{-0.102}^{+0.097}$ & $0.297_{-0.017}^{+0.018}$\\
\hline
\multirow{2}{*}{\shortstack{DESI,\\ Pantheon+}}
& Bayesian                       & $-0.88_{-0.11}^{+0.13}$ & $-0.27_{-0.87}^{+0.88}$ & $0.304_{-0.050}^{+0.029}$ & $0.305_{-0.015}^{+0.016}$ \\
& Frequentist                    & $-0.88_{-0.12}^{+0.13}$ & $-0.25_{-0.87}^{+0.94}$ & $0.307_{-0.066}^{+0.026}$ & $0.305 \pm 0.016$\\
\hline
\multirow{2}{*}{\shortstack{DESI,\\ DES Y5}}
& Bayesian                       & $-0.78_{-0.13}^{+0.14}$ & $-0.75_{-0.89}^{+0.90}$ & $0.321_{-0.033}^{+0.025}$ & $0.311_{-0.015}^{+0.016}$\\
& Frequentist                    & $-0.78_{-0.13}^{+0.15}$ & $-0.73_{-0.91}^{+0.90}$ & $0.321_{-0.036}^{+0.025}$ & $0.310 \pm 0.016$\\

\end{tabular}
\end{ruledtabular}
\caption{\textbf{95\%} Bayesian credible intervals (mean $\pm 2\sigma$) and frequentist confidence intervals (MLE $\pm 2\sigma$) for $w_0$, $w_a$ {\rm and } $\Omega_\mathrm{m}$ within the CPL model, and $\Omega_\mathrm{m}$ within $\Lambda$CDM. Also at the $2\sigma$-level, we find good agreement between the Bayesian and frequentist intervals.
}
\label{tab:constraints_2sigma}
\end{table*}


\section{Profile likelihoods without CMB data}
\label{app:noCMB}

\begin{figure*}
    \centering
    \includegraphics[width=0.95\linewidth]{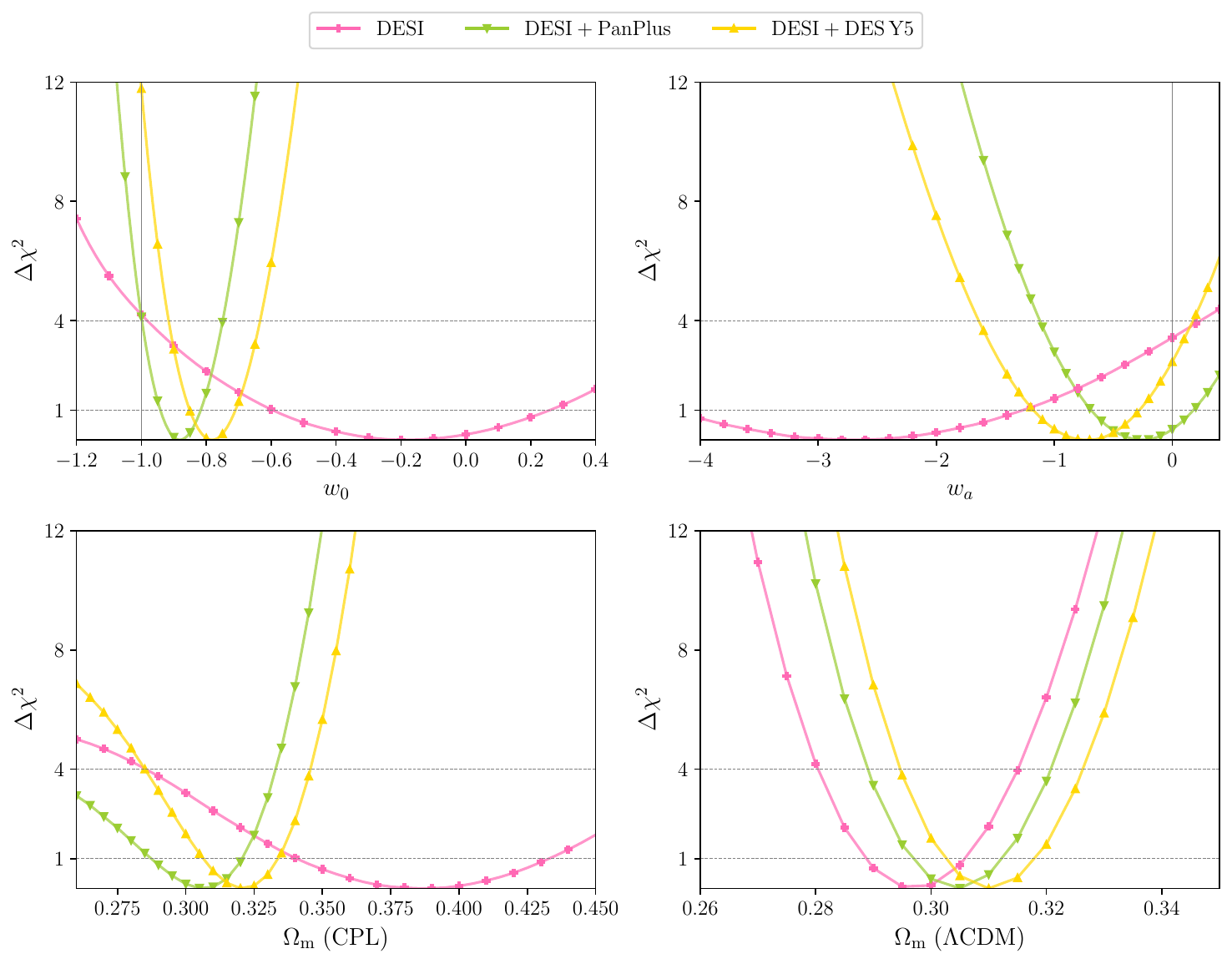}
    \caption{\textit{Top:} Profile likelihoods in $w_0$ (left) and $w_a$ (right) for dataset combinations without the CMB as indicated in the legend. 
    \textit{Bottom:} Profile likelihoods in $\Omega_\mathrm{m}$ within CPL (left) and $\Lambda$CDM (right).
    }
    \label{fig:PLs_noCMB}
\end{figure*}

Fig.~\ref{fig:PLs_noCMB} shows the profile likelihoods of $w_0$, $w_a$ and $\Omega_\mathrm{m}$ within CPL and $\Lambda$CDM for the dataset combinations excluding the CMB, i.e.\ DESI alone and DESI combined with SNe. These profiles are also parabolic within $1\sigma$ but significantly deviate from a parabola at $2\sigma$.

\section{Contributions to the \texorpdfstring{$\chi^2$ of $w_0\, w_a$}{chi2 of w0, wa}}
\label{app:chi2contr}

\begin{figure*}
    \centering
    \includegraphics[width=0.95\linewidth]{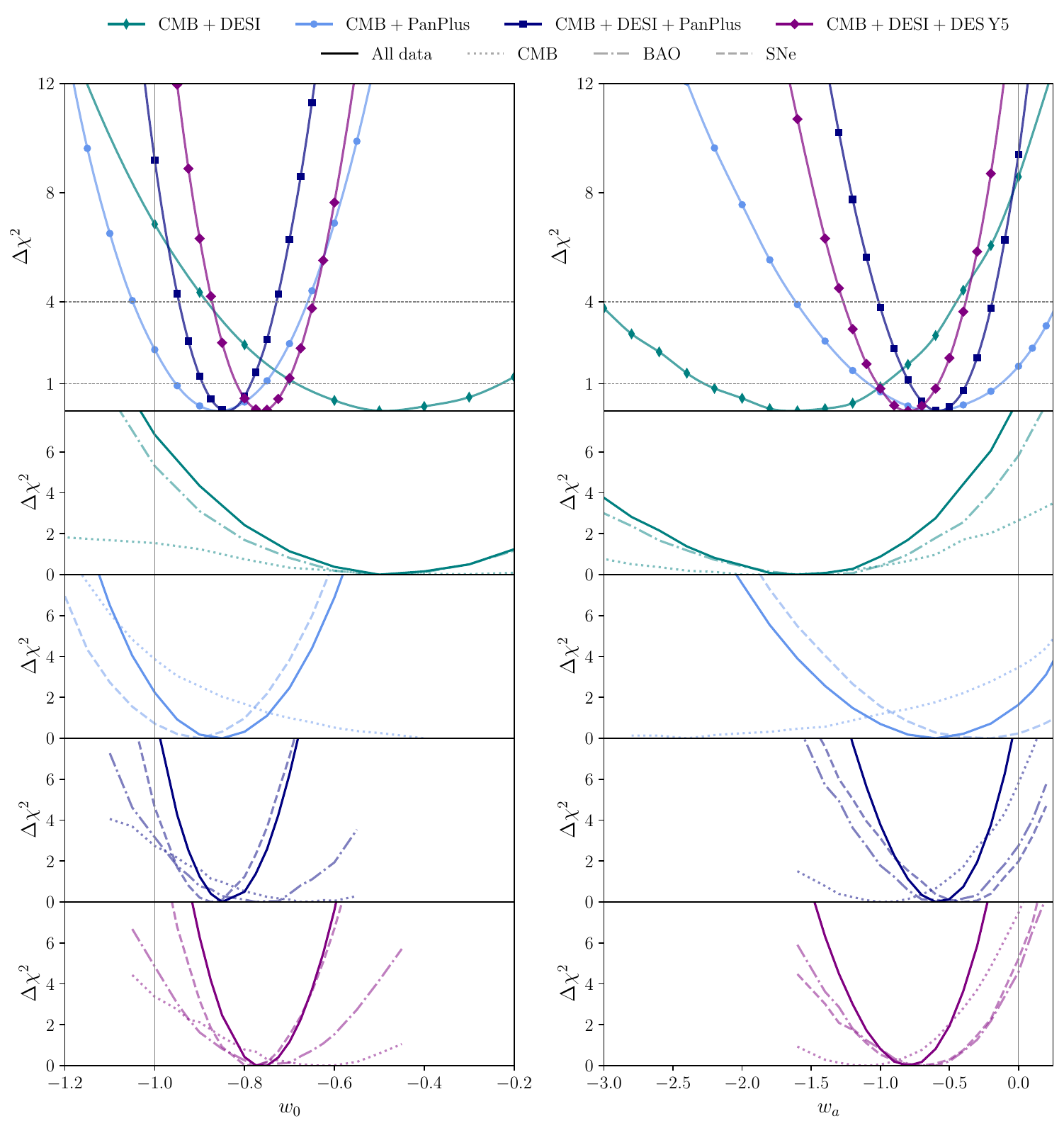}
    \caption{\textit{Top:} Profile likelihoods in $w_0$ (left) and $w_a$ (right); the same as \cref{fig:PLs_w0wa}. 
    \textit{Other panels:} For each dataset combination, indicated by the colors in the legend, the respective panel shows the total $\Delta\chi^2$ (solid) along with the contributions from the individual datasets: CMB (dotted), DESI BAO (dash-dotted) and SNe (dashed), c.f.\ Eq.~\ref{eq:chi2_contributions}. Each curve is individually adjusted by subtracting out its minimum $\chi^2$.
    }
    \label{fig:PLs_w0wa_contr}
\end{figure*}

\cref{fig:PLs_w0wa_contr} shows the profile likelihoods in $w_0$ and $w_a$, along with the contributions to the $\chi^2$ from individual datasets. As expected, the preference for $w_0 > -1$ and $w_a < 0$ is dominated by DESI and Pantheon+ or DES Y5 data. Nevertheless, CMB consisting of \Planck\ primary CMB combined with \Planck+ACT lensing shows a weak preference for the same quadrant of parameter space, as already pointed out in \cite{Planck:2018vyg}, pushing the constraints to slightly higher $w_0$ and slightly lower $w_a$.

\cref{tab:chi2} shows the contributions to the $\chi^2$ from the individual datasets at the MLE. 
Going from \LCDM\ to the $w_0 w_a$ model improves the fit for all dataset combinations, with two additional degrees of freedom. While the improvement $\Delta\chi^2 = -2.5$ for CMB and Pantheon+ data is negligible, the improvement $\Delta\chi^2 \approx-10$ for CMB, DESI (and Pantheon+) is significant, corresponding to almost $3\sigma$ for a $\chi^2$ distribution with two degrees of freedom. For CMB, DESI and DES Y5 data, we find the largest improvement of fit $\Delta\chi^2 \approx -20$, corresponding to over $4\sigma$. These $\chi^2$ differences are consistent with \cite{DESI:2025zgx}. For each dataset combination, the fit to the individual datasets, CMB, BAO and SNe, respectively, is improved in $w_0w_a$ compared to \LCDM. Thus, all datasets considered in this work contribute to the preference for $w_0 > -1$, $w_a <0$.

\begin{table*}
\begin{ruledtabular}
\begin{tabular}{lc|llll}
\multirow{2}{*}{Data combination} & \multirow{2}{*}{Model} & \multicolumn{3}{c}{$\chi^2$ ($\Delta \chi^2$)} \\
                                  &                         & Total & CMB & BAO & SNe \\
\hline
\multirow{2}{*}{CMB, DESI\ } 
& \LCDM & 2800.2 & 2788.4 & 11.7 & - \\
& $w_0w_a$ & 2790.5 (-9.7) & 2783.4 (-5.0) & 7.0 (-4.7) & - \\
\hline
\multirow{2}{*}{\shortstack{CMB, Pantheon+}} 
& \LCDM & 4195.6 & 2784.9 & - & 1410.6 \\
& $w_0w_a$ & 4193.1 (-2.5) & 2783.5 (-1.4) & - & 1409.6 (-1.1) \\
\hline
\multirow{2}{*}{\shortstack{CMB, DESI,\ Pantheon+}} 
& \LCDM & 4213.1 & 2787.6 & 12.8 & 1412.7 \\
& $w_0w_a$ & 4203.2 (-9.9) & 2784.0 (-3.6) & 9.5 (-3.2) & 1409.7 (-3.0) \\
\hline
\multirow{2}{*}{\shortstack{CMB, DESI, DES Y5}}
& \LCDM & 4449.6 & 2787.1 & 13.6 & 1648.8 \\
& $w_0w_a$ & 4429.9 (-19.6) & 2783.7 (-3.4) & 8.7 (-4.9) & 1637.5 (-11.3) \\
\hline
\multirow{2}{*}{\shortstack{DESI}}
& \LCDM & 10.3 & - & 10.3 & - \\
& $w_0w_a$ & 5.6 (-4.7) &  - & 5.6 (-4.7) &  -\\
\hline
\multirow{2}{*}{\shortstack{DESI, Pantheon+}}
& \LCDM & 1423.3 & - & 10.9 & 1412.4 \\
& $w_0w_a$ & 1420.9 (-2.4) &  - & 10.0 (-0.9) &  1410.9 (-1.5) \\
\hline
\multirow{2}{*}{\shortstack{DESI, DES Y5}}
& \LCDM & 1659.0 & - & 12.4 & 1646.5 \\
& $w_0w_a$ & 1645.4 (-13.6) &  - & 8.0 (-4.4) &  1637.5 (-9.0) \\
\end{tabular}
\end{ruledtabular}
\caption{
The $\chi^2$ breakdown across datasets at the best-fit cosmologies for each model. In parentheses, we also show the improvement $\Delta\chi^2 = \chi^2_{w_0w_a} - \chi^2_{\Lambda {\rm CDM}}$ in goodness of fit under $w_0w_a$ relative to \LCDM for each dataset. 
}
\label{tab:chi2}
\end{table*}

\bibliography{main}

\end{document}